\documentclass[aps,pre,twocolumn,showpacs,10pt]{revtex4-1}

\usepackage{graphicx}
\usepackage{dcolumn}
\usepackage{bm}
\usepackage{amsmath}
\usepackage{amssymb}
\usepackage[utf8x]{inputenc}
\usepackage[T1]{fontenc}
\usepackage[%
  colorlinks=true,%
  linkcolor=blue,%
  citecolor=blue,%
  urlcolor=blue%
]{hyperref}
\usepackage{tikz}
\usepackage{pgfplots}
\usetikzlibrary{pgfplots.groupplots}

\pgfplotsset{compat=newest}
\newcommand{\ii}{\mathrm{i}}
\newcommand{\ee}{\mathrm{e}}

\usepackage{txfonts}
\DeclareMathAlphabet{\mathcal}{OMS}{cmsy}{m}{n}

\begin{document}

\title{Semiclassical bifurcations and topological phase transitions in a one-dimensional lattice of coupled Lipkin-Meshkov-Glick models}

\author{A. V. Sorokin\textsuperscript{1}}
\email{a.sorokin@mailbox.tu-berlin.de}
\author{M. Aparicio Alcalde\textsuperscript{1}}
\author{V. M. Bastidas\textsuperscript{1,2}}
\author{G. Engelhardt\textsuperscript{1}}
\author{D. G. Angelakis\textsuperscript{2,3}}
\author{T. Brandes\textsuperscript{1}}
\affiliation{\textsuperscript{1}%
	Institut für Theoretische Physik, Technische Universität Berlin, Hardenbergstr. 36, 10623 Berlin, Germany%
}
\affiliation{\textsuperscript{2}%
	Centre for Quantum Technologies, National University of Singapore, 3 Science Drive 2, Singapore 117543%
}
\affiliation{\textsuperscript{3}%
	School of Electronic and Computer Engineering, Technical University of Crete, Chania, Crete, Greece, 73100%
}

\begin{abstract}
	In this work we study a one-dimensional lattice of Lipkin-Meshkov-Glick models with alternating couplings between nearest-neighbors sites, which resembles the Su-Schrieffer-Heeger model. Typical properties of the underlying models are present in our semiclassical-topological hybrid system, allowing us to investigate an interplay between semiclassical bifurcations at mean-field level and topological phases. Our results show that bifurcations of the energy landscape lead to diverse ordered quantum phases. Furthermore, the study of the quantum fluctuations around the mean field solution reveals the existence of nontrivial topological phases. These are characterized by the emergence of localized states at the edges of a chain with open boundary conditions.
\end{abstract}


\maketitle

\section{Introduction}
\label{sec:intro}

Rather recently, an enormous interest has arisen in the investigation of topological effects in diverse fields. Condensed matter has been a fruitful arena to study topological insulators~\cite{Hasan2010,Bernevig2006,Fu2007,Fu2007a,Hsieh2008,Konig2007,Thouless1982} and topological properties of superconductors~\cite{Kitaev2001,Alicea2012}. There are proposals for the experimental realizations of topological states by using a quantum wire with Rashba spin-orbit interaction, a combination of Zeeman splitting and proximity-induced $s$-wave superconductivity~\cite{2010vonOppen,Lutchyn2010} or semiconductor quantum dots coupled to superconducting grains~\cite{2012NatCo...3E.964S}. Recent experiments found signatures of topologically protected states in nanowires coupled to superconductors~\cite{2012Kouwenhoven} and in ferromagnetic atomic chains on the surface of a superconducting lead~\cite{2014Yazdani}. The experimental observation and manipulation of these states will have a strong impact on quantum technologies and quantum information processing. Topologically-protected excitations of a topological superconductor are so-called Majorana fermions which exhibit nonabelian statistics. This can be used to perform braiding operations, which are essential for the implementation of fault-tolerant quantum computation~\cite{Liu2013,Alicea2012}.

The interest in topological states of matter is not only restricted to condensed matter systems. Currently, these systems can be simulated by using ultracold gases in optical lattices~\cite{Lewenstein2007,LiuTop2013}.
There are also proposals to realize exotic topological states in quantum optics by exploiting light-matter interactions~\cite{PeanoPhysRevX.5.031011}, such as fractional quantum Hall effect in arrays of coupled cavities~\cite{2008Angelakis,2012Carusotto}, topological superradiance~\cite{2015Guang-Can}, and optical realizations of the Jackiw-Rebbi model~\cite{2014Angelakis}. In optics, it has been shown that metamaterials are optical analogs of topological insulators~\cite{2013Shvets,PeanoPhysRevX.5.031011} and they can be used to perform topological pumping~\cite{2016ArxivLee}. Recently, topological states have been observed in photonic quantum walks~\cite{Kitagawa2010a,Kitagawa2012}, photonic quasicrystals~\cite{2013Silberberg}, and self-localized states in photonic topological insulators~\cite{2013Segev}. These developments have such a strong influence in material science that currently, one can design phononic topological metamaterials where elastic waves inherit topological features~\cite{2015Zheng,susstrunk2015observation}. Chiral spin-wave edge modes in dipolar magnetic systems~\cite{Shindou2013,Shindou2013a,Shindou2014} are another example.

In this work, we are inspired by yet another example of topological behavior in quantum physics: the Su-Schrieffer-Heeger ({\sc ssh}) model of polyacetylene, which is a polymer with alternating bonds~\cite{1979-Su-PRL,*1980-Su-PRB}. This model exhibits a topological quantum phase transition~\cite{ChenPhysRevB.89.085111}. In the topological phase, there are topologically protected edge states, which are localized at the ends of an open chain. Besides its natural appearing in polyacetylene, the {\sc ssh} model has been realized by using a chain of dielectric microwave resonators~\cite{2015Schomerus} and in a system of cold atoms in an optical potential~\cite{Lohse2015}.

In our paper, we investigate a semiclassical-topological hybrid model, which enables us to investigate the influence of a semiclassical bifurcation on topological properties. To this end, we consider a one-dimensional chain of coupled Lipkin-Meshkov-Glick ({\sc lmg}) models with alternating couplings resembling the original {\sc ssh} model~\cite{1979-Su-PRL,*1980-Su-PRB}, as shown in Fig.~\ref{fig:Chain}. 
The {\sc lmg} model describes an ensemble of $M$ two-level systems with all-to-all anisotropic interactions~\cite{1965-Lipkin-NuclPhys,1965-Meshkov-NuclPhys,1965-Glick-NuclPhys}. In particular, the {\sc lmg} model exhibits a semiclassical bifurcation, which has been observed experimentally~\cite{2010-Zibold-PRL}. Although the bifurcation takes place at mean-field level, there are quantum signatures of this effect in the dynamics of the system~\cite{BrandesPhysRevA.91.013631,2014-Muessel-PRL}.

Among our findings, we show that for a system of coupled {\sc lmg} models in the semiclassical limit $j\gg 1$ one can define an energy surface at mean-field level. In particular, the semiclassical bifurcation of an isolated {\sc lmg} model gives rise to a quantum phase transition ({\sc qpt}) of a spatial symmetry-breaking type. Similar to an isolated {\sc lmg} model, here, the form of the quasienergy landscape determines quantum fluctuations. As a consequence, on the quantum level, the semiclassical bifurcation is accompanied by the emergence of squeezing visible in the fluctuations around the mean field.  The effect of squeezing on topological properties of the system has been recently investigated~\cite{PeanoTop2016,2015-Georg-PRA,Furukawa2015}. In contrast, the main benefit of our approach is that we provide a microscopic and analytically tractable  model to study the origin of the squeezing terms and their consequences. Moreover, the combination of topological features and of a classical bifurcation transition in our model allows to investigate the interplay of these two effects.

The structure of this article is as follows. In Sec.~\ref{sec:model}, we introduce the model and discuss the semiclassical limit and the quantum fluctuations around the semiclassical trajectories. In Sec.~\ref{sec:bulk}, we consider a chain with periodic boundary conditions. After the discrete Fourier transformation, we define the Bogoliubov Hamiltonian and the excitation spectrum, which enables us to calculate topological properties. This section is followed by the discussion of edge states in a finite chain with open boundary conditions presented in Sec.~\ref{sec:finite}. Conclusive remarks are presented in Sec.~\ref{sec:conclusion}.

\section{Model}\label{sec:model}

\subsection{Hamiltonian and symmetries}

\begin{figure}
	\begin{tikzpicture}[x=1pt, y=1pt, >=stealth]
    \fill[fill=black!10](-12,-24) rectangle +(64,6pc);
    \draw[line width=2pt](0,0) -- node[above]{$\kappa_1$} (32,18);
    \draw(32,18) -- node[above]{$\kappa_2$} (64,0);
    \draw[line width=2pt](64,0) -- (96,18);
    \draw[gray,dashed](96,18) -- (128,0);
    \draw[line width=2pt](128,0) -- (160,18);
    \begin{scope}[line width=2pt, blue, ->]
        \draw(0,0) -- +(5,24);
        \draw(32,18) -- +(5,24);
        \draw(64,0) -- +(-5,24);
        \draw(96,18) -- +(-5,24);
        \draw(128,0) -- +(5,24);
        \draw(160,18) -- +(5,24);
    \end{scope}
    \begin{scope}[ball color=orange]
        \shade(0,0) circle(5);
        \shade(32,18) circle(5);
        \shade(64,0) circle(5);
        \shade(96,18) circle(5);
        \shade(128,0) circle(5);
        \shade(160,18) circle(5);
    \end{scope}
    \path(0,-14) node {$A1$}
         (32,-14) node {$B1$}
         (64,-14) node {$A2$}
         (96,-14) node {$B2$}
         (112,-14) node {$\cdots$}
         (128,-14) node {$AN$}
         (160,-14) node {$BN$};
    \begin{scope}[draw=black!33]
        \draw(160,18) circle(7);
        \draw(166.546,18.422) -- (180.562,20.832);
        \draw(210,24) circle(30);
    \end{scope}
    \draw(204,45) -- (216,45);
    \draw(204,33) -- (216,33);
    \draw(192,18) -- (204,18);
    \draw(192,6) -- (204,6);
    \draw(216,18) -- (228,18);
    \draw(216,6) -- (228,6);
    \draw[black!66,<->](219,6) -- node[right]{$\omega_0$} (219,18);
    \draw[black!66,<->](201,6) -- node[left]{$\omega_0$} (201,18);
    \draw[black!66,<->](213,33) -- node[left]{$\omega_0$} (213,45);
    \draw[green!66,thick,<->](204,12) -- (216,12);
    \draw[green!66,thick,<->](198,18) -- (207,33);
    \draw[green!66,thick,<->](222,18) -- (213,33);
\end{tikzpicture}
	\caption{\label{fig:Chain}(Color online) Chain of {\sc lmg} models with alternating couplings $\kappa_1$ and $\kappa_2$. Grey-shaded elementary unit cell contains two nodes ($A$ and $B$). Each node is characterized by its pseudospin $\bm{J}_{Pl}$ and parameters $\omega_0$ and $\gamma$.}
\end{figure}
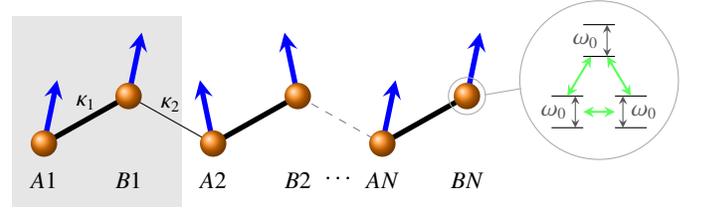

In this work, we study quantum-critical and topological aspects of a one-dimensional lattice of coupled many-body systems. We consider a chain of $2N$ sites, which is built up in a configuration of alternating coupling strengths resembling the Su-Schrieffer-Heeger ({\sc ssh}) model (see Fig.~\ref{fig:Chain}). The unit cell of the lattice consists of two sites labeled $A$ and $B$ (see Fig.~\ref{fig:Chain}). The local dynamics of the sites are governed by the  {\sc lmg} model~\cite{1965-Lipkin-NuclPhys,*1965-Meshkov-NuclPhys,*1965-Glick-NuclPhys}, which describes a set of $M$ two-level systems with all to all coupling. Its Hamiltonian for a given site $P\in\{A,B\}$ in the $l$th unit cell with $ l\in\{1,2,\ldots,N\}$ reads
\begin{equation}
	\hat{H}_{Pl} = \omega_0 J^{z}_{Pl}-\frac{\gamma}{2j}(J_{Pl}^x)^{2},
	\label{eq:LMG}
\end{equation}
where $ J_{Pl}^{\mu}= \frac 12\sum_{i=1}^M \sigma_{Pl}^{\mu,i}$ with $\mu,\nu,\rho\in\{x,y,z\}$ being collective angular momentum operators and $\sigma_{Pl}^{\mu,i}$ denoting common Pauli matrices. Collective operators fulfill $[J_{Pl}^{\mu},J_{P'l'}^{\nu}]=\ii\delta_{PP'}\delta_{ll'}\varepsilon^{\mu\nu\rho}J_{Pl}^{\rho}$, where $\varepsilon^{\mu\nu\rho}$ is the Levi-Civita symbol. The parameter $\omega_0$ denotes the level splitting of the two-level systems, $\gamma$ is the self-interaction strength. Throughout the article we restrict ourselves to the subspace of maximal total spin quantum number $j=M/2$. For example, in the realization of the {\sc lmg} model in a Bose-Einstein condensate, the system is naturally restricted to this subspace~\cite{2010-Zibold-PRL,PhysRevE.86.012101}.

 In order to study quantum fluctuations around a given mean-field configuration, we consider the bosonic representation of the angular-momentum algebra, known as the Holstein-Primakoff transformations~\cite{1940-Holstein-PhysRev}
\begin{gather}
	J_{Pl}^z = b_{Pl}^\dagger b_{Pl}^{\phantom\dagger}-j,\nonumber\\
	J_{Pl}^+ = b_{Pl}^\dagger\sqrt{2j-b_{Pl}^\dagger b_{Pl}^{\phantom\dagger}},\qquad
	J_{Pl}^- = \sqrt{2j-b_{Pl}^\dagger b_{Pl}^{\phantom\dagger}}\;b_{Pl}^{\phantom\dagger},
	\label{eq:HP}
\end{gather}
where $b_{Pl}$ and $b_{Pl}^\dagger$ are bosonic operators satisfying the commutation relations $[b_{Pl}^{\phantom\dagger}, b_{P'l'}^\dagger]=\delta_{PP'}\delta_{ll'}$ and $[b_{Pl}^{\phantom\dagger}, b_{P'l'}^{\phantom\dagger}]=[b_{Pl}^\dagger, b_{P'l'}^\dagger]=0$.

Besides the  dynamics governed by the local Hamiltonian~\eqref{eq:LMG}, we consider alternating nearest-neighbors interactions, which can be of ferromagnetic or antiferromagnetic type. The total Hamiltonian of the lattice reads
\begin{equation}
	\hat{H}=\sum_{l=1}^N (\hat{H}_{Al}+\hat{H}_{Bl}) - \frac1{2j}\sum_{l=1}^N(\kappa_1 J_{Al}^+ J_{Bl}^-+\kappa_2 J_{Bl}^+ J_{Al+1}^-+\text{h.c.}),
	\label{eq:SSH-LMG}
\end{equation}
where $\kappa_1$ and $\kappa_2$ are the intracell and intercell nearest-neighbor coupling constants, respectively. In order to simplify expressions, the total Hamiltonian and all the coupling constants are normalized by $\omega_0$ hereafter.

A complete understanding of the symmetries of the system enables one to find conserved quantities, which simplifies the study of the physical problem. In the thermodynamic limit, some of these symmetries can be spontaneously broken, marking the onset of a quantum phase transition. However, changes of phase can also occur without symmetry breaking. Such phase transitions are referred to as topological quantum phase transitions.

Let us consider the operator
\begin{equation}
	\hat{N}=\sum_{\substack{l=1 \\ P=A,B}}^N\left(J_{Pl}^z+j\right)=\sum_{\substack{l=1 \\ P=A,B}}^N b_{Pl}^\dagger b_{Pl}^{\phantom\dagger},
	\label{eq:GlobalJz}
\end{equation}
which is the $z$-component of the total angular momentum (up to an additive constant). In the bosonic representation, $\hat{N}$ is the number operator, which counts the total number of excitations in the $J^z$ basis.

The Hamiltonian~\eqref{eq:SSH-LMG} commutes with $\hat{N}$ only for $\gamma=0$, meaning that  the total number of excitations is conserved. Moreover,  the model remains invariant under an arbitrary global rotation $\hat{R}_\theta^z=\ee^{\ii\theta\hat{N}}$ around the $z$-axis, which corresponds to a $\mathrm{SU}(1)$ continuous symmetry. In the case $\gamma\neq 0$, however, the Hamiltonian~\eqref{eq:SSH-LMG} lacks this symmetry, but is still invariant under a discrete parity symmetry $\hat{R}_\pi^z=\ee^{\ii\pi\hat{N}}$.

\subsection{Mean field and  fluctuations}
\label{sec:GS-QH}

Throughout the following sections, we are interested in the properties of the system~\eqref{eq:SSH-LMG} in the semiclassical limit $j\gg 1$. In this limit, each isolated site of the lattice---which corresponds to an {\sc lmg} model---exhibits a semiclassical bifurcation at mean-field level~\cite{2007-Ribeiro-PRL}. Besides the semiclassical limit, we also consider the thermodynamic limit $N\rightarrow \infty$, when the symmetries of the system can be spontaneously broken and the collective $J_z$ operator in Eq.~\eqref{eq:GlobalJz} can have finite values. 

\begin{figure}
	\raggedleft
	\definecolor{colPhase}{rgb}{1., 0.933333, 0.666667}
\begin{tikzpicture}[x=1pc, y=1pc]
    \fill[colPhase](0,3) -- (0,4) -- (4,4) -- (4,0) -- (3,0) -- cycle;
    \fill[colPhase](0,-3) -- (0,-3.5) -- (-3.5,-3.5) -- (-3.5,0) -- (-3,0) -- cycle;
    \fill[colPhase](0,3) -- (0,4) -- (-3.5,4) -- (-3.5,0) -- (-3,0) -- cycle;
    \fill[colPhase](0,-3) -- (0,-3.5) -- (4,-3.5) -- (4,0) -- (3,0) -- cycle;
    \draw[gray!50](-8pt,-9pt) rectangle (4,4);
    \draw(3,0) -- (0,3) -- (-3,0) -- (0,-3) -- cycle;
    \path(10pt,-10pt) node{I} (2.5,2.5) node{II} (-2.5,2.5) node{IV}
         (-2.5,-2.5) node{III} (2.5,-2.5) node{V};
    \draw[->](-3.5,0) -- (4,0) node[anchor=south east, inner sep=2pt]{$\kappa_2$};
    \draw[->](0,-3.5) -- (0,4) node[anchor=north west, inner sep=2pt]{$\kappa_1$};
    \filldraw[fill=gray, draw=blue](-3,1) circle(1.5pt);
    \filldraw[fill=gray!50, draw=blue](-2,1) circle(1.5pt);
    \filldraw[fill=white, draw=blue](-0.5,1) circle(1.5pt);
    \node[fill=white, draw=red, inner sep=1.5pt] at (0.5,1) {};
    \node[fill=gray!50, draw=red, inner sep=1.5pt] at (2,1) {};
    \node[fill=gray, draw=red, inner sep=1.5pt] at (3,1) {};
    \path(3,0) node[anchor=north] {\scriptsize 0.5}
         (0,3) node[anchor=east] {\scriptsize 0.5}
         (0,0) node[anchor=north east] {\scriptsize 0};
    \draw(142pt,0) -- node[shape=rectangle,sloped,fill=white,inner sep=1]{\scriptsize 0.0} ++(-40pt,40pt)
        -- ++(-40pt,-40pt) -- ++(40pt,-40pt) -- cycle;
    \draw(134pt,0) -- ++(-32pt,32pt) -- node[shape=rectangle,sloped,fill=white,inner sep=1]{\scriptsize 0.2}
           ++(-32pt,-32pt) -- ++(32pt,-32pt) -- cycle;
    \draw(126pt,0) -- node[shape=rectangle,sloped,fill=white,inner sep=1]{\scriptsize 0.4} ++(-24pt,24pt)
        -- ++(-24pt,-24pt) -- ++(24pt,-24pt) -- cycle;
    \draw(118pt,0) -- ++(-16pt,16pt) -- node[shape=rectangle,sloped,fill=white,inner sep=1]{\scriptsize 0.6}
           ++(-16pt,-16pt) -- ++(16pt,-16pt) -- cycle;
    \draw[->](5,0) -- (12.5,0) node[above]{$\kappa_2$};
    \draw[->](8.5,-3.5) -- (8.5,4) node[right]{$\kappa_1$};
    \draw(104pt,40pt) -- (100pt,40pt) node[color=black!50,left]{1};
    \draw(142pt,2pt) -- (142pt,-2pt) node[color=black!50,below]{1};
    \path(4,-3.5) node[anchor=south east, inner sep=0]{(a)}
         (150pt,-3.5) node[anchor=south east, inner sep=0]{(b)};
\end{tikzpicture}\vspace{5pt}
	\begin{tikzpicture}
    \pgfplotstableread{bands.dat}\tabBands
    \begin{groupplot}[group style={%
            group size=2 by 3,%
            xlabels at=edge bottom,%
            xticklabels at=edge bottom,%
            ylabels at=edge left,%
            yticklabels at=edge left,%
            vertical sep=0pt, horizontal sep=0pt},%
        footnotesize, no markers,%
        width=0.6\columnwidth, height=7pc,%
        xmin=-3.1416, xmax=3.1416, ymin=0, ymax=0.8,%
        ytick align=outside,xtick align=outside,%
        tickpos=left, xlabel=Wavenumber $k$,%
        xtick={-3.1416, 0}, ytick={0, 0.2, 0.4, 0.6},%
        xticklabels={$-\pi$, $0$},%
        minor xtick={0}, grid=minor]
        \nextgroupplot[extra y ticks={0.8}]
            \addplot[blue] table[x index=0, y index=1]{\tabBands};
            \addplot[blue] table[x index=0, y index=2]{\tabBands};
            \filldraw[fill=gray, draw=blue](-1.5416,0.2452) circle(1.5pt);
        \nextgroupplot
            \addplot[red] table[x index=0, y index=11]{\tabBands};
            \addplot[red] table[x index=0, y index=12]{\tabBands};
            \node[fill=gray, draw=red, inner sep=1.5pt] at (-1.5416,0.2415) {};
        \nextgroupplot[ylabel=Excitation energy $\varepsilon_\lambda$]
            \addplot[blue] table[x index=0, y index=3]{\tabBands};
            \addplot[blue] table[x index=0, y index=4]{\tabBands};
            \filldraw[fill=gray!50, draw=blue](-1.5416,0.1386) circle(1.5pt);
        \nextgroupplot
            \addplot[red] table[x index=0, y index=9]{\tabBands};
            \addplot[red] table[x index=0, y index=10]{\tabBands};
            \node[fill=gray!50, draw=red, inner sep=1.5pt] at (-1.5416,0.1331) {};
        \nextgroupplot
            \addplot[blue] table[x index=0, y index=5]{\tabBands};
            \addplot[blue] table[x index=0, y index=6]{\tabBands};
            \filldraw[fill=white, draw=blue](-1.5416,0.2573) circle(1.5pt);
        \nextgroupplot[extra x ticks={3.1416}, extra x tick labels={$\pi$}]
            \addplot[red] table[x index=0, y index=7]{\tabBands};
            \addplot[red] table[x index=0, y index=8]{\tabBands};
            \node[fill=white, draw=red, inner sep=1.5pt] at (-1.5416,0.2546) {};
            \path(axis cs:3.1416,0) node[anchor=south east, inner sep=1pt]{(c)};
    \end{groupplot}
\end{tikzpicture}
	\caption{\label{fig:Phases}(Color online) (a)~Bifurcation diagram for the system. Disordered phase I and four symmetry-broken phases II--V. The bifurcation from one to two degenerate ground states occurs at the phase boundaries of I. The grey box is for the future reference. (b)~Dependence of the phase boundary on the value of $\gamma$ (labels of the contours). For $\gamma\geqslant 1$ only symmetry-broken phases are present. (c)~Bulk band structure for particular points of the phase diagram depicted in panel a as calculated from Eqs.~\eqref{eq:Ek-I} and~\eqref{eq:Ek-II}. Parameter values used in panel c are: $\gamma=0.5$, $\kappa_1=0.15$ and (from top to bottom) $\kappa_2=\pm0.5$, $\pm0.35$, $\pm0.1$. The left and right columns of panel c correspond to the negative and positive values of the intercell coupling $\kappa_2$, respectively.}
\end{figure}
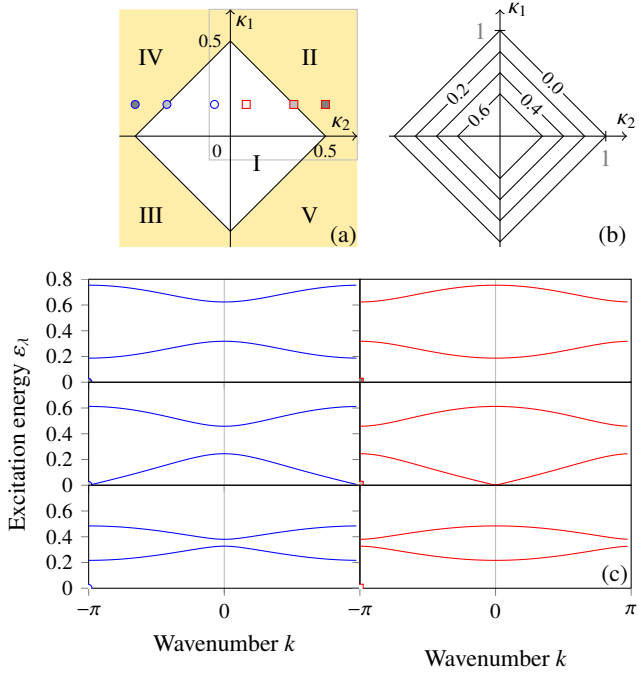

In the limit $j\gg 1$, we perform a mean-field analysis of the system  to understand semiclassical features of the model such as bifurcations~\cite{BuchleitnerPhysRevB.93.155153}. In order  to study quantum signatures of these semiclassical bifurcations, we must calculate the quantum fluctuations around the semiclassical trajectories. With this aim in mind, we introduce a time-dependent local displacement operator $\hat{D}_{Pl}[\alpha_{Pl}(t)]=\exp\left[\sqrt{j}\left(\alpha_{Pl}^*(t) b_{Pl}^{\phantom\dagger}-\alpha_{Pl}^{\phantom *}(t) b_{Pl}^{\dagger}\right)\right]$, where $\alpha_{Pl}(t)$ is a time-dependent mean field. To simplify the notation, we drop the explicit time dependence of the mean fields in the remainder of the article.
By using the displacement operator, we define a new set of displaced bosonic operators~\cite{1969-Glauber-PhysRev}
\begin{equation}
	d_{Pl}=\hat{D}_{Pl}^\dagger(\alpha_{Pl}) b_{Pl}^{\phantom\dagger} \hat{D}_{Pl}^{\phantom\dagger}(\alpha_{Pl})=b_{Pl}-\alpha_{Pl}\sqrt{j},
	\label{eq:D}
\end{equation}
which describe the fluctuations around the mean fields $\alpha_{Pl}\sqrt{j}$. By first applying the Holstein-Primakoff transformation~\eqref{eq:HP} to the Hamiltonian~\eqref{eq:SSH-LMG} and then applying the displacement operation~\eqref{eq:D} to the Schrödinger equation, the latter finally reads
\begin{equation}
	\ii\partial_{t}|\Psi_{\bm{\alpha}}(t)\rangle=\hat{H}_{\bm{\alpha}}|\Psi_{\bm{\alpha}}(t)\rangle,
\end{equation}
where
\begin{gather*}
	|\Psi_{\bm{\alpha}}(t)\rangle = \hat{D}^\dagger(\bm{\alpha})|\Psi(t)\rangle,\quad
	\hat{D}(\bm{\alpha})=\bigotimes_{Pl}\hat{D}_{Pl}(\alpha_{Pl}),\\
	\hat{H}_{\bm{\alpha}}=\hat{D}^\dagger(\bm{\alpha})\left(\hat{H}-\ii\partial_{t}\right) \hat{D}(\bm{\alpha}),\\
	\hat{D}^{\dagger}(\bm{\alpha})\partial_t \hat{D}(\bm{\alpha})=%
	\frac{j}{2}(\dot{\bm{\alpha}}\bm{\alpha}^*-\bm{\alpha}\dot{\bm{\alpha}}^*)+\sqrt{j}(\dot{\bm{\alpha}}\bm{d}^{\dagger}-\dot{\bm{\alpha}}^*\bm{d})
\end{gather*}
and $\bm{\alpha}=(\alpha_{A1},\alpha_{B1},\ldots,\alpha_{AN},\alpha_{BN})$.

To identify the dominant mean-field contribution and its fluctuations in the Hamiltonian $\hat{H}_{\bm{\alpha}}$, we  expand the latter as a power series in $1/\sqrt{j}$. In the semiclassical limit $j\gg 1$, the series can be truncated~\cite{2014-Sorokin-PRE} as
\begin{equation}
	\hat{H}_{\bm{\alpha}} = j L(\bm{\alpha})+
	\sqrt{j} \hat{H}_L(\bm{d},\bm{\alpha})+
	\hat{H}_Q(\bm{d},\bm{\alpha}).
	\label{eq:H-GLQ}
\end{equation}
The terms $\hat{H}_L(\bm{d},\bm{\alpha})$ and $\hat{H}_Q(\bm{d},\bm{\alpha})$ are linear and quadratic in the bosonic operators $d_{Pl}$, respectively~\cite{2011-Hayn-PRA,*2012-Hayn-PRA}. The term proportional to $j$ corresponds to the Lagrangian $L(\bm{\alpha})=-\frac{\ii}{2}(\dot{\bm{\alpha}}\bm{\alpha}^*-\bm{\alpha}\dot{\bm{\alpha}}^*)+E(\bm{\alpha})$, where
\begin{multline}
	E(\bm\alpha)=-2N+\sum_{l=1}^N(\alpha_{Al}^*\alpha_{Al}^{\phantom *}+\alpha_{Bl}^*\alpha_{Bl}^{\phantom *})\\
	-\frac{\gamma}{8}\sum_{l=1}^N\left[(2-\alpha_{Al}^*\alpha_{Al}^{\phantom *})(\alpha_{Al}^*+\alpha_{Al}^{\phantom *})^2+(2-\alpha_{Bl}^*\alpha_{Bl}^{\phantom *})(\alpha_{Bl}^*+\alpha_{Bl}^{\phantom *})^2\right]\\
	-\frac{\kappa_1}{2}\sum_{l=1}^N\sqrt{2-\alpha_{Al}^*\alpha_{Al}^{\phantom *}}\sqrt{2-\alpha_{Bl}^*\alpha_{Bl}^{\phantom *}}(\alpha_{Al}^*\alpha_{Bl}^{\phantom *}+\alpha_{Bl}^*\alpha_{Al}^{\phantom *})\\
	-\frac{\kappa_2}{2}\sum_{l=1}^N\sqrt{2-\alpha_{Al+1}^*\alpha_{Al+1}^{\phantom *}}\sqrt{2-\alpha_{Bl}^*\alpha_{Bl}^{\phantom *}}(\alpha_{Al+1}^*\alpha_{Bl}^{\phantom *}+\alpha_{Bl}^*\alpha_{Al+1}^{\phantom *})
	\label{eq:Eg}
\end{multline}
is the classical Hamilton function. The  transformation that we have used can be interpreted as a transformation into a co-moving frame~\cite{ArmourPhysRevLett.104.053601,HammererPhysRevX.4.011015}. The center of the frame is given by the mean fields $\bm{\alpha}(t)$. We require that the linear terms $\hat{H}_L(\bm{d},\bm{\alpha})$ in Eq.~\eqref{eq:H-GLQ} vanish. This is the case, when the mean fields satisfy the classical Hamilton equations 
\begin{equation}
	\left\{\dot{\alpha}_{Pl}^*=\ii\frac{\partial}{\partial\alpha_{Pl}} E(\bm\alpha),\quad\dot{\alpha}_{Pl}=-\ii\frac{\partial}{\partial\alpha_{Pl}^*} E(\bm\alpha)\right\}_{\substack{l=1,\ldots,N \\ P=A,B}}.
	\label{eq:minConditions}
\end{equation}
In doing so, the mean fields $\bm{\alpha}(t)$ follow the trajectory of the corresponding semiclassical system.

Once we solve this set of equations, we are prepared to calculate the quantum fluctuations governed by the quadratic Hamiltonian $\hat{H}_Q(\bm{d},\bm{\alpha})$ from Eq.~\eqref{eq:H-GLQ}. In general, for a given semiclassical trajectory $\bm{\alpha}(t)$, the quadratic Hamiltonian is time dependent. Of particular interest are stationary semiclassical trajectories that do not evolve in time, i.e., critical points of the function $E(\bm\alpha)$ satisfying the conditions $\dot{\alpha}_{Pl}^*=\dot{\alpha}_{Pl}=0$. In this article, we focus precisely on this kind of trajectories, as we intend to investigate the  ground-state properties.
 In the case of the bifurcation of the energy landscape, even an exponentially-scaling number of these can appear~\cite{2014-Sorokin-PRE}.
Interestingly, Gaussian quantum fluctuations are determined by the Hessian, and, as a consequence, they depend on the local geometry of the Hamiltonian function $E(\bm\alpha)$ in phase space.

\subsection{Ground state}

In this section, we explain how to calculate the ground state of the system and its properties. In doing so, we do not specify boundary conditions of the chain. Thus, the following explanations are valid for both periodic and open boundary conditions, which we consider in more detail in Secs.~\ref{sec:bulk} and~\ref{sec:finite}, respectively.

As we previously discussed, at the minima $\bm{\alpha}_m$ of the energy landscape $E(\bm{\alpha})$ the linear terms in Eq.~\eqref{eq:H-GLQ} vanish, i.e., $\hat{H}_L(\bm{d},\bm{\alpha}_m)=0$~\cite{2011-Hayn-PRA}. Consequently, the quadratic part $\hat{H}_Q(\bm{d},\bm{\alpha}_m)$ of the Hamiltonian~\eqref{eq:H-GLQ} describes the fluctuations on top of the ground state. They are responsible for the low-lying  excitations in the system. To simplify the notation, we drop the explicit dependence of the quadratic Hamiltonian on the mean field and bosonic operators, i.e., $\hat{H}_Q=\hat{H}_Q(\bm{d},\bm{\alpha}_m)$, which is equal to
 \begin{multline}
 	\hat{H}_Q = \\
 	\sum_{l=1}^N\Big[c^{(1)}(\alpha_{Al},\alpha_{Bl},\alpha_{Bl-1}) d_{Al}^{\dagger}d_{Al}^{\phantom\dagger} + c^{(1)}(\alpha_{Bl},\alpha_{Al},\alpha_{Al+1}) d_{Bl}^{\dagger}d_{Bl}^{\phantom *}\Big] \\
 	+\sum_{l=1}^N\Big[c^{(2)}(\alpha_{Al},\alpha_{Bl},\alpha_{Bl-1}) d_{Al}^{2} + c^{(2)}(\alpha_{Bl},\alpha_{Al},\alpha_{Al-1}) d_{Bl}^{2}+\text{h.c.}\Big] \\
 	+\sum_{l=1}^N\Big[\kappa_1 c^{(3)}(\alpha_{Al},\alpha_{Bl}^{\phantom *}) d_{Al}^{\dagger}d_{Bl}^{\phantom *} + \kappa_2 c^{(3)}(\alpha_{Al+1},\alpha_{Bl}) d_{Al+1}^{\dagger}d_{Bl}+\text{h.c.}\Big] \\
 	+\sum_{l=1}^N\Big[\kappa_1 c^{(4)}(\alpha_{Al},\alpha_{Bl}) d_{Al}d_{Bl} + \kappa_2 c^{(4)}(\alpha_{Al+1},\alpha_{Bl}) d_{Al+1}d_{Bl}+\text{h.c.}\Big].
 	\label{eq:HQ}
 \end{multline}
This expression is even valid for arbitrary displacement in Eq.~\eqref{eq:D}. In Appendix~\ref{AppendixA}, we provide the explicit form of the coefficients. We note, that by treating the system classically by means of a Landau-Lifshitz equations, one can obtain a formally equivalent result by linearizing the classical equations of motion~\cite{Shindou2013,Shindou2013a,Shindou2014}.

In  this work, $\kappa_1$ and $\kappa_2$ are assumed to take any real values. Obviously, the Hamiltonian~\eqref{eq:SSH-LMG} is not invariant under the interchange of the signs of the couplings
\begin{equation}
	\kappa_1 \to (-1)^{u}\kappa_1 \quad\text{and}\quad
	\kappa_2 \to (-1)^{v}\kappa_2,
	\label{eq:trafoKappa}
\end{equation}
where $u,v\in\{0,1\}$. However, there is a local gauge transformation, which compensates this change of signs:
\begin{align}
	J_{Al}^\eta &\rightarrow (-1)^{(u+v)l} J_{Al}^\eta, &
	J_{Al}^z &\rightarrow J_{Al}^z, \nonumber \\
	J_{Bl}^\eta &\rightarrow (-1)^{(u+v)l+u} J_{Bl}^\eta, &
	J_{Bl}^z &\rightarrow  J_{Bl}^z
	\label{eq:trafoAngMomentum}
\end{align}
for $\eta=x,y$ or, correspondingly,
\begin{align}
	b_{Al} &\rightarrow (-1)^{(u+v)l} b_{Al}, \nonumber \\
	b_{Bl} &\rightarrow (-1)^{(u+v)l+u} b_{Bl}.
	\label{eq:trafoHPoperators}
\end{align}

By virtue of these transformations, we can confine our calculations to the region with positive couplings only and from the results reconstruct all the properties of the other regions. For example, to consider the Hamiltonian~\eqref{eq:SSH-LMG}  with $\kappa_1>0$ and $\kappa_2<0$, we transform it using Eq.~\eqref{eq:trafoKappa} for $u=0$ and $v=1$. In doing so, we obtain a similar Hamiltonian, but with positive couplings. Then we  use the  results of the following sections and transform them back using Eq.~\eqref{eq:trafoAngMomentum} again. Using an analogous procedure and transformations~\eqref{eq:trafoKappa} and~\eqref{eq:trafoAngMomentum}, we can also analyze the other cases with different values of the couplings $\kappa_{1,2}$.

It is possible to apply the above analysis to study the expanded Hamiltonian~\eqref{eq:H-GLQ}. In this sense, for any signs of $\kappa_{1,2}$, Eq.~\eqref{eq:D} together with transformations~\eqref{eq:trafoKappa} and~\eqref{eq:trafoHPoperators} defines new mean values $\alpha'_{Pl}$ and displaced operators $d'_{Pl}$.
Thereby, the mean fields $\alpha_{Al}$ and $\alpha_{bl}$ are transformed the same way as $b_{Al}$ and $b_{Bl}$, respectively. This allows us  to effectively minimize the ground-state energy for all $\kappa_{1,2}$ using the positive coupling mean-field minima $\alpha_{Pl}=\alpha_{m\,Pl}$. It is not difficult to also prove that coefficients $c^{(i)}$ of the quadratic Hamiltonian~\eqref{eq:HQ} remain intact for different couplings with $|\kappa_{1,2}|$ fixed, thus $\hat{H}_Q$ depends only on signs of couplings $\kappa_{1,2}$.

The approach described in this section allows us to study the ground state and its quantum fluctuations for chains with both periodic and open boundary conditions. The next two sections deal with these two types of boundary conditions separately.

\section{Bulk properties}
\label{sec:bulk}

In this section, we study the translationally invariant system. This is crucial for understanding the effects in a corresponding finite-size system: It is a fundamental characteristic of topological insulators, that the properties of a translationally invariant system are reflected by the presence or absence of localized midgap states close the boundary of a finite-size system with open boundary conditions. This relation is denoted as the bulk--boundary relation~\cite{Hasan2010}.

The ground-state energy is obtained by minimizing $E(\bm{\alpha})$, given in Eq.~\eqref{eq:Eg}. For this model, the minima $\alpha_{Pl}$ are summarized in Tab.~\ref{tab:A}, where mean values $\alpha_{Pl}$  define five quantum phases~I--V in Fig.~\ref{fig:Phases}a. Spatially uniform minima are obtained analytically for the phase~II. In the previous section we explained how to obtain the minima of phases~III--V from the phase~II using Eqs.~\eqref{eq:trafoKappa} and~\eqref{eq:trafoHPoperators}. The ground-state energy in phase~I is 
\begin{equation}
	E(\bm{\alpha}_m)=-2N,
\end{equation}
and in phases II--V one finds that
\begin{equation}
	E(\bm{\alpha}_m)=-\left[2-\frac{(1-\gamma-|\kappa_1|-|\kappa_2|)^2}{\gamma+|\kappa_1|+|\kappa_2|}\right]N.
\end{equation}
As a consequence, the ground-state energy is not analytic at the phase transition and has a discontinuity in the second derivatives with respect to $\kappa_1$ or $\kappa_2$. To this end, the system exhibits a quantum phase transition of second order as the isolated {\sc lmg} model~\cite{2008-Ribeiro-PRE}.  

\begin{table}
	\caption{\label{tab:A}Mean values of bosonic operators $b_{Pl}$ in respective phases. The shorthand $a=\ee^{\ii\phi}\sqrt{1-1/(\gamma+|\kappa_1|+|\kappa_2|)}$ is used. $\phi$ is arbitrary for $\gamma=0$ and assumes values $0$ or $\pi$ for $\gamma\neq 0$.}
	\begin{ruledtabular}
		\begin{tabular}{cccccc}
			Phase: & I & II & III & IV & V \\
			\hline
			$\alpha_{Al}$ & 0 & $a$ & $a$ & $(-1)^l a$ & $(-1)^l a$ \\
			$\alpha_{Bl}$ & 0 & $a$ & $-a$ & $-(-1)^l a$ & $(-1)^l a$ \\
		\end{tabular}
	\end{ruledtabular}
\end{table}

\begin{table}
	\caption{\label{tab:C}Coefficients $c^{(i)}$ ($i=1,\ldots,4$) defining $\mathcal{H}_Q$ in each of the phases. The shorthand $\zeta=\gamma+|\kappa_1|+|\kappa_2|$ is used.}
	\begin{ruledtabular}
		\begin{tabular}{ccccc}
			Phase & $c^{(1)}$ & $c^{(2)}$ & $c^{(3)}$ & $c^{(4)}$ \\ \hline
			I & $1-\gamma/2$ & $-\gamma/4$ & $-1$ & $0$ \\
			II--V & $\frac{5\zeta^3+2\zeta^2+\zeta-4\gamma}{4\zeta(\zeta+1)}$ & $\frac{3\zeta^3-2\zeta^2-\zeta-4\gamma}{8\zeta(\zeta+1)}$ & $-\frac{\zeta^2+2\zeta+5}{4\zeta(\zeta+1)}$ & $\frac{\zeta^2+2\zeta-3}{4\zeta(\zeta+1)}$
		\end{tabular}
	\end{ruledtabular}
\end{table}

Since periodic boundary conditions are taken, coefficients of the quadratic Hamiltonian~\eqref{eq:HQ} are position-independent and can be computed at minima points. This allows us to obtain a simplified expression
\begin{align}
	\hat{H}_Q &= \sum_{l=1}^{N}c^{(1)}(d_{Al}^\dagger d_{Al}^{\phantom\dagger}+ d_{Bl}^\dagger d_{Bl}^{\phantom\dagger}) + \sum_{l=1}^{N}c^{(2)}\left( d_{Al}^2 + d_{Bl}^2+\text{h.c.}\right) \nonumber\\
		&\quad +\sum_{l=1}^{N}\kappa_1\left(c^{(3)}d_{Al}^\dagger d_{Bl}^{\phantom\dagger}+ c^{(4)}d_{Al}^{\phantom\dagger} d_{Bl}^{\phantom\dagger} + \text{h.c.}\right)\nonumber\\
		&\quad +\sum_{l=1}^{N}\kappa_2\left(c^{(3)} d_{Bl}^\dagger d_{Al+1}^{\phantom\dagger} + c^{(4)} d_{Bl}^{\phantom\dagger} d_{Al+1}^{\phantom\dagger} + \text{h.c.}\right).
\label{eq:SimpleHQ}
\end{align}
The values of the coefficients are summarized in Tab.~\ref{tab:C}. We proceed by mapping the bosonic operators $d_{Pl}$ onto the reciprocal space by the Fourier transforms
\begin{equation}
	d_{Pl}=\frac1{\sqrt{N}}\sum_{k}{\tilde{d}_{Pk} \ee^{\ii k l}},
\label{eq:DFT}
\end{equation}
where $k=\frac{2\pi n}{N}$ with $n=\left\{-\frac{N'}{2},-\frac{N'}{2}+1,\ldots,\frac{N'}{2}-1\right\}$ and $N'=N$ when $N$ is even and $N'=N-1$ when $N$ is odd. $\tilde{d}_{Pl}$ are bosonic operators, satisfying $[\tilde{d}_{Pl}^{\phantom\dagger}, \tilde{d}_{P'l'}^\dagger]=\delta_{PP'}\delta_{ll'}$ and $[\tilde{d}_{Pl}^{\phantom\dagger}, \tilde{d}_{P'l'}^{\phantom\dagger}]=[\tilde{d}_{Pl}^\dagger, \tilde{d}_{P'l'}^\dagger]=0$. After the discrete Fourier transformation, the Hamiltonian $\hat{H}_Q$ can be written as
\begin{equation*}
	\hat{H}_Q=\frac{1}{2}\sum_{k}\left(\begin{array}{cc}
		\tilde{\bm d}_k^\dagger & \tilde{\bm d}_{-k}^T
	\end{array}\right)
	\bm{H}_k
	\left(\begin{array}{c}
		\tilde{\bm d}_k \\ (\tilde{\bm d}^\dagger_{-k})^T
	\end{array}\right),
\end{equation*}
where $\tilde{\bm d}_k=(\tilde{d}_{Ak}\;\tilde{d}_{Bk})^T$ and $\tilde{\bm{d}}^\dagger_k=(\tilde{d}^\dagger_{Ak} \;\tilde{d}^\dagger_{Bk})$. Correspondingly, we obtain the Bogoliubov Hamiltonian 
\begin{equation*}
	\bm{H}_k=\left(\begin{array}{cc}
		\bm{A}_k & \bm{B}_k\\
		\bm{B}_k & \bm{A}_k
	\end{array}\right),
	\label{eq:BogHk}
\end{equation*}
where
\begin{align*}
	\bm{A}_k &= \left(\begin{array}{cc}
		c^{(1)} & c^{(3)}(\kappa_1+\kappa_2\ee^{-\ii k})\\
		c^{(3)}(\kappa_1+\kappa_2\ee^{\ii k}) & c^{(1)}
	\end{array}\right),\\
	\bm{B}_k &= \left(\begin{array}{cc}
		2c^{(2)} & c^{(4)}(\kappa_1+\kappa_2\ee^{-\ii k})\\
		c^{(4)}(\kappa_1+\kappa_2\ee^{\ii k}) & 2c^{(2)}
	\end{array}\right).
\end{align*}

The matrix $\bm{H}_k$ can be diagonalized following the Bogoliubov theory as  explained in Ref.~\onlinecite{2012-Kawaguchi-PhysRep}. To this end, we define a $4\times 4$ transformation matrix $\bm{T}_k$ such that $(\tilde{\bm d}_k^\dagger\;\tilde{\bm d}_{-k}^T)=(\tilde{\bm r}_k^\dagger\;\tilde{\bm r}_{-k}^T)\,\bm{T}_k^{\dagger}$. Because  the new operators $\tilde{\bm r}_k$ must fulfill bosonic commutation relations, the matrix $\bm{T}_k$ must satisfy $\bm{T}_k\bm{\sigma}_z\bm{T}_k^\dagger=\bm{\sigma}_z$, where $\bm{\sigma}_z=\mathrm{diag}(1,1,-1,-1)$. The diagonalized quadratic Hamiltonian finally reads
\begin{equation}
	\hat{H}_Q=\sum_{k}\tilde{\bm r}_k^\dagger\bm{\varepsilon}_k^{\phantom\dagger}\tilde{\bm r}_k^{\phantom\dagger},
\end{equation}
where $\bm{\varepsilon}_k$ is a $2\times 2$ diagonal matrix of excitation energies. The matrix $\bm{T}_k$ and $\bm{\varepsilon}_k$ can be found by solving the eigenvalue equation
\begin{equation}
	\bm{\sigma}_z\bm{H}_k\,\bm{T}_k=\bm{T}_k\left(\begin{array}{cc}
		\bm{\varepsilon}_k & \bm{0}\\
		\bm{0} & -\bm{\varepsilon}_{-k}
	\end{array}\right),
\end{equation}
where the components of the diagonal matrices $\bm{\varepsilon}_{\pm k}$ are the eigenvalues of the matrix $\bm{\sigma}_z\bm{H}_k$ and the respective eigenvectors are the columns of the matrix $\bm{T}_k$.

In the following, we use  the notation $(\bm{\varepsilon}_{k})_{\lambda,\lambda}=\varepsilon_\lambda(k)$. Due the fact that there are two basis states ($A$ and $B$) within a unit cell, dispersion relation has two bands, with labels $\lambda=1,2$, respectively. Choosing $\varepsilon_1(k)\leqslant\varepsilon_2(k)$ for each of the phases~I--V, we obtain dispersion relations. In the disordered phase~I we find
\begin{multline}
	\varepsilon_\lambda^{(\text{I})}(k)=\sqrt{1+(-1)^\lambda\sqrt{\kappa_1^2+\kappa_2^2+2\kappa_1\kappa_2\cos k}}\\
	\times\sqrt{1-\gamma+(-1)^\lambda\sqrt{\kappa_1^2+\kappa_2^2+2\kappa_1\kappa_2\cos k}}
\label{eq:Ek-I}
\end{multline}
and in the ordered phases II--V it is
\begin{multline}
	\varepsilon_{\lambda}^{(\text{II--V})}(k)=\frac1\zeta \Bigg[\zeta(\zeta^3-\gamma)+\kappa_1^2+\kappa_2^2+2\kappa_1\kappa_2\cos k\\
	+(-1)^\lambda(\zeta^3+\zeta-\gamma)\sqrt{\kappa_1^2+\kappa_2^2+2\kappa_1\kappa_2\cos k}\Bigg]^{\frac12},
	\label{eq:Ek-II}
\end{multline}
where $\zeta=\gamma+|\kappa_1|+|\kappa_2|$. From Eqs.~\eqref{eq:Ek-I} and~\eqref{eq:Ek-II} one can see that the simultaneous change of signs of both couplings $\kappa_{1,2}\rightarrow -\kappa_{1,2}$ does not affect the band structure, whereas the change of sign of just one of the couplings shifts the band structure by $\pi$. Important characteristics of dispersion relations could be seen from Fig.~\ref{fig:Phases}c.

\subsection{Topological properties}
\label{sec:topology}

Our model is motivated by the {\sc ssh} model, which can exhibit a topological band structure. As we demonstrate in the following, our strongly interacting system inherits this topological feature. A central question of the current work is how the interactions and the bifurcation influence topological signatures of the system. In particular, in Sec.~\ref{sec:finite} we show that the main effect of the interaction is that it can obscure the existence of topologically protected edge states. 

An essential feature of systems with underlying topology is that there appear symmetry-protected edge states, which are strongly localized close to the boundaries of a finite-size system with open boundary conditions. Moreover, the celebrated bulk-boundary relation states that the number of these edge states is given by a quantized topological invariant describing a translationally invariant system. For a one-dimensional topological insulator without interactions, the topological quantum number is a winding number. References~\onlinecite{2015-Georg-PRA} and~\onlinecite{Shindou2013} show how to generalize this topological invariant to bosonic Bogoliubov systems. We call such a generalized invariant {\em symplectic polarization} in the following. It reads
\begin{equation}
	P_s = \frac\ii{2\pi}\int_{\text BZ}\left[\bm{\sigma}_z \bm{T}_k^\dagger\bm{\sigma}_z\left(\frac{d}{dk}\bm{T}_k\right)\right]_{1,1} dk
	\label{eq:SimplecticP}
\end{equation}
and is a real-valued quantity. It is not hard to see that our system obeys an inversion symmetry relation 
\begin{gather*}
 	\bm{\tau}_x\bm{H}_k\bm{\tau}_x=\bm{H}_{-k},\\
	\bm{\tau}_x=\textstyle\left(\begin{array}{cc}
		1&0\\0&1
	\end{array}\right)\otimes
	\textstyle\left(\begin{array}{cc}
		0&1\\1&0
	\end{array}\right).
\end{gather*}

Reference~\onlinecite{2015-Georg-PRA} proves that this is a sufficient condition   for the quantization of the symplectic polarization. More precisely, the symplectic polarization can take values $P_s= 0,\frac12, 1, \frac32,\ldots$ Importantly, as a topological invariant of a noninteracting system, $P_s$ can not be changed by a smooth transformation of the system that does not close the gap between the bands $\lambda=1$ and $\lambda=2$. Thus, a change of topology is always marked by a gap closing.
 
\begin{figure*}
	\includegraphics[width=0.9\textwidth]{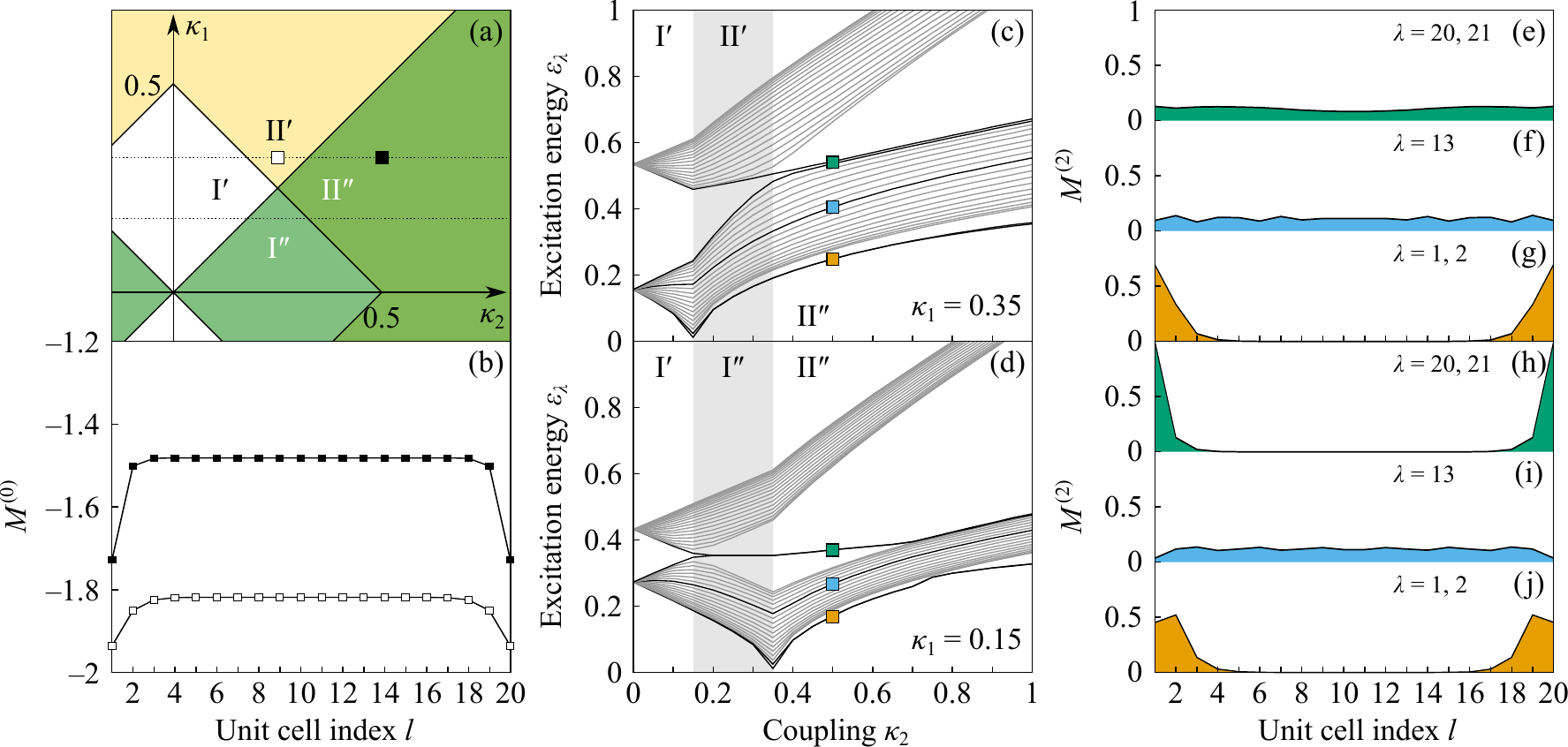}
	\caption{\label{fig:TopoPhases}(Color online) (a)~Topological and quantum phases of the system (cf.~grey box in Fig.~\ref{fig:Phases}a). Couplings $\kappa_{1,2}$ are assumed positive. Primed phases are topologically trivial while doubly primed phases (green-shaded) are non-trivial. (b)~Macroscopic magnetization depending on position in the chain for $\gamma=0.5$, $\kappa_1=0.35$, $\kappa_2=0.25$ (empty squares) and $\kappa_2=0.5$ (filled squares). (c,~d)~Excitation energies for the open chain with $N=20$, $\gamma=0.5$ and $0\leqslant\kappa_2\leqslant 1$ along the lines $\kappa_1=0.35$ and $\kappa_1=0.15$ respectively. (e--j)~Microscopic fluctuations of the magnetization depending on the position in the chain and the number of the band. $\gamma=0.5$, $\kappa_2=0.5$. For panels e--g, $\kappa_1=0.35$; for panels h--j, $\kappa_1=0.15$. Each of the plots corresponds to respectively color-coded points in panels~c and d.}
\end{figure*}

The gap closing  takes place  at $k=\pi$ when $\kappa_1 \kappa_2>0$ and at $k=0$ otherwise [as derived from Eq.~\eqref{eq:Ek-I} for phase~I and from Eq.~\eqref{eq:Ek-II} for phases~II--V]. In all cases, the resulting boundary is found to be $|\kappa_1|=|\kappa_2$|. By explicitly evaluating $P_s$ we  verify that phases I$'$ and II$'$ (see Fig.~\ref{fig:TopoPhases}a) are topologically \emph{trivial} with $P_s=0$ , while phases I$''$ and II$''$ are topologically \emph{non-trivial} with $P_s=\frac12$ (green-shaded in Fig.~\ref{fig:TopoPhases}a). For other regions of the phase diagram, the picture is symmetric by reflection with respect to the $\kappa_1$ and $\kappa_2$ axes.

\section{Finite chain with open boundaries}
\label{sec:finite}

\subsection{Ground-state energy and excitation energies}
\label{sec:excitations}

Though the topological quantum number $P_s$ is defined for the translationally-invariant Hamiltonian, it is hard to infer the topology of a bosonic system by considering bulk observables. However, we recall that $P_s$ can predict edge states localized at the boundary of a finite-size chain with open boundary conditions. Thereby, the number of edge states is given by $2P_s$.

As we consider open boundary conditions, we can not transform the Hamiltonian into momentum space and we have to perform the calculations in position space. Due to the size of such systems, it is not possible to diagonalize Bogoliubov matrices analytically, so we are restricted to numerical calculations.

As a consequence of the boundary effects,  we cannot adopt the uniform ansatz ($\alpha_{Pl}=\alpha$). First, we have to find the  mean field configuration  which  minimize the ground-state energy~\eqref{eq:Eg}. For that we use the method of conjugate gradients~\cite{1952-Hestenes-JRes-CG,1964-Fletcher-ComputJ-CG} with the uniform ansatz as its zeroth approximation. Due to the fact that only a small number of $\alpha_{Pl}$ at the edges of the chain are to be optimized as well as the spatial symmetry of the system, the algorithm works robustly and converges to reliable results for all sets of parameters we are interested in. 

After minimizing the  mean-field ground-state energy, we follow the same approach as before. As finite chains with open boundaries lack translational invariance, we cannot perform transformations into Fourier space, and we use the Hamiltonian~\eqref{eq:HQ}. Because of the non-uniformity of the mean fields, the coefficients $c^{(i)}$ are no longer constant and have to be recalculated for each $\alpha_{Pl}$ separately. We can rewrite the Hamiltonian~\eqref{eq:HQ} as a quadratic form
\begin{equation}
	\hat{H}_Q=\frac{1}{2}\left(\begin{array}{cc}
		\bm{d}^\dagger & \bm{d}^T
	\end{array}\right) \bm{H} \left(\begin{array}{c}
		\bm{d} \\ \bm{d}^{\dagger\,T}
	\end{array}\right),
	\label{eq:finHQ}
\end{equation}
where we have defined $\bm{d}=\left(d_{A1},\,d_{B1},\ldots,\,d_{AN},\,d_{BN}\right)^T$ and $\bm{d}^{\dagger}=\left(d_{A1}^{\dagger},\,d_{B1}^{\dagger},\ldots,\,d_{AN}^{\dagger},\,d_{BN}^{\dagger}\right)$. In addition, the matrix representation of the quadratic form has the block structure
\begin{equation}
	\bm{H}=\left(\begin{array}{cc}
		\bm{A} & \bm{B} \\ \bm{B} & \bm{A}
	\end{array}\right).
	\label{eq:finitHQ}
\end{equation}
We now perform the symplectic diagonalization~\cite{2012-Kawaguchi-PhysRep} of the $4N\times 4N$ matrix $\bm{H}$, where $N$ is the number of unit cells in the chain. The diagonalization procedure essentially works as described in Sec.~\ref{sec:bulk}, but with $ \bm{H}_k$ replaced by $\bm{H}$. Accordingly, one has to adjust the dimensions of $\bm{\sigma}_z=\sigma_z \otimes \mathbf 1_N$, where $\mathbf 1_N$ denotes the identity matrix of dimension $N$.

Let us now look at the excitation energies in Figs.~\ref{fig:TopoPhases}c,d. We fix $\gamma=0.5$, one of the couplings ($\kappa_1=0.15$ and $0.35$ for c and d cases respectively) and vary $\kappa_2$. The two-band structure resembles the one of the {\sc ssh} model: In the topologically non-trivial phase the lowest state of the upper band separates and meets in the band gap with the highest state of the lower band.

In contrast to the {\sc ssh} model, another pair of states separates beneath the lower band in the quantum phase~II (Figs.~\ref{fig:TopoPhases}c,d; marked with orange squares). These states are related to the parameter $\gamma$, which is responsible for spontaneous symmetry breaking. We emphasize that these localized states are not related to the symplectic polarization~\eqref{eq:SimplecticP}, as the latter predicts only the edge states that are energetically located within the gap between bands with $\lambda=1$ and $\lambda=2$. These sub-band states are the consequence of the deviations of the mean fields $\alpha_{Pl}$ close to the boundary from the mean fields within the bulk. This effectively creates an impurity close to the boundary which results in a localized eigenstates according to ordinary scattering theory~\cite{Economou1984}.

\subsection{Magnetization}
\label{sec:magnetization}

In order to make connection to potential experimental realizations, one has to consider measurable observables. As we have a (pseudo-)spin system, the obvious choice of such an observable is the spin component $\langle J_{Pl}^z\rangle$ or, as the system is constructed from unit cells, the magnetization $M_l = \langle J_{Al}^z + J_{Bl}^z\rangle$. To calculate the latter, we first map spin operators onto respective bosonic operators, obtaining
\begin{align}
	M_l &= \langle d_{Al}^\dagger d_{Al}^{\phantom\dagger}\rangle + \langle d_{Bl}^\dagger d_{Bl}^{\phantom\dagger}\rangle \nonumber\\
		&\quad +\left(\alpha_{Al}^*\langle d_{Al}^{\phantom\dagger}\rangle + \alpha_{Al}^{\phantom *}\langle d_{Al}^\dagger\rangle + \alpha_{Bl}^*\langle d_{Bl}^{\phantom\dagger}\rangle + \alpha_{Bl}^{\phantom *}\langle d_{Bl}^\dagger\rangle\right)\sqrt{j}\nonumber\\
		&\quad + (\alpha_{Al}^*\alpha_{Al}^{\phantom\dagger} + \alpha_{Bl}^*\alpha_{Bl}^{\phantom\dagger} - 2)j\nonumber\\
		&= M_l^{(0)} + M_l^{(1)}\sqrt{j} + M_l^{(2)} j.
\end{align}
This expression contains terms of orders $O(j^0)$, $O(j^{1/2})$ and $O(j^1)$. The $O(j^0)$ terms constitute the macroscopic magnetization $M^{(0)}_l$, which depends on the mean fields $\alpha_{Pl}$ only, so it remains the same for any (reasonably low) number of excitations of any of the normal modes of the system. The $O(j^{1/2})$ terms are zero, as for a harmonic oscillator $\langle d_{Al}^{(\dagger)}\rangle=0$.

The $O(j^0)$ terms constitute microscopic fluctuations $M^{(2)}_l$ on top of the macroscopic magnetization. To calculate these, we first have to chose a suitable state in which to take the expectation value. A reasonable choice would be $|\lambda\rangle = |0_1 0_2 \cdots 1_\lambda \cdots 0_{2N}\rangle$, i.e. the state in which there is only one excitation in the $\lambda$th normal mode, but all the other modes are empty. This choice allows us to investigate, how the magnetization differs in different edge and bulk states of the system. Taking into account that the transformation matrix
\[
	\bm{T}=\begin{pmatrix}
		\bm{U} & \bm{V}^* \\ \bm{V} & \bm{U}^*
	\end{pmatrix}
\]
has the block form, we obtain
\begin{align}
	\langle\lambda|d_{\bm{s}}^\dagger d_{\bm{s}}^{\phantom\dagger}|\lambda\rangle &=
		\langle\lambda|\sum_{n,m}\left(V_{\bm{s}n}r_n+U_{\bm{s}n}^*r_n^\dagger\right)\left(U_{\bm{s}m}r_m+V_{\bm{s}m}^*r_m^\dagger\right)|\lambda\rangle\nonumber\\
		&=\langle\lambda|\sum_{n,m}\left(V_{\bm{s}n}V_{\bm{s}m}^*r_nr_m^\dagger + U_{\bm{s}n}^*U_{\bm{s}m}r_n^\dagger r_m\right)|\lambda\rangle\nonumber\\
		&=\langle\lambda|\sum_n\left(|V_{\bm{s}n}|^2r_nr_n^\dagger + |U_{\bm{s}n}|^2r_n^\dagger r_n\right)|\lambda\rangle\nonumber\\
		&=\langle\lambda|\sum_n\left[(|V_{\bm{s}n}|^2+|U_{\bm{s}n}|^2)r_n^\dagger r_n + |V_{\bm{s}n}|^2\right]|\lambda\rangle\nonumber\\
		&=|V_{\bm{s}\lambda}|^2+|U_{\bm{s}\lambda}|^2 + \sum_n |V_{\bm{s}n}|^2,
\end{align}
where $d_{\bm{s}}=\sum_{n}(U_{\bm{s}n}r_n+V_{\bm{s}n}^*r_n^\dagger)$ is a Bogoliubov transformation, $U_{\bm{s}n}$ and $V_{\bm{s}m}$ are the entries of the matrices $\bm{U}$ and $\bm{V}$, respectively, and $m,n\in\{1,\ldots,2N\}$. To simplify the expressions, we have introduced the notation $\bm{s}=Ps$, where $P\in\{A,B\}$ denotes the bosonic species and $s\in\{1,\dots,N\}$ denotes the unit cell.

Fig.~\ref{fig:TopoPhases}b depicts the macroscopic part of the magnetization in phases II$'$ and II$''$. We observe that the chain may be separated in the edge part that contains two unit cells and the translationally-invariant bulk part. Edge effects are more pronounced in the topologically nontrivial phase. In the phase I we have  $\alpha_l=0$ for any parameter values both in the bulk and at the edges, thus the macroscopic magnetization is also constant throughout the phase: $M^{(0)}=-2$.

\subsection{Edge states}

As the macroscopic magnetization is a  ground-state property, it is not possible to get any information about edge states out of it. In order to compare edge states and bulk states, we have to go beyond the ground state and look for signatures in the fluctuations of the magnetization $M^{(2)}_{l,\lambda}$. To this end, we performed calculations in the topologically nontrivial symmetry-broken phase II$''$---as it is of greatest interest---computing $M^{(2)}_{l,\lambda}$ for the lower edge states ($\lambda=1,2$), for a state within a band ($\lambda=13$) and for the edge states in the band gap ($\lambda=19,20$) using two different sets of parameters (see the caption of Fig.~\ref{fig:TopoPhases}). The results are plotted in Figs.~\ref{fig:TopoPhases}e--j. As usual eigenstates, the edges states inherit the symmetry of their underlying Hamiltonian. The Hamiltonian obeys a reflection symmetry, so it also holds for expectation values of the magnetisation in the edge states in Fig.~\ref{fig:TopoPhases}e--j. More precisely, there is always one edge with even parity and one with odd parity under spatial inversion. Their energy difference vanishes exponentially with the chain length. Therefore, here we find a pair of edge states within the band gap.

From these plots one can draw several conclusions. Firstly, as was expected, the microscopic magnetization in the bulk is more or less uniformly delocalized and rather low for both sets of parameters (Figs.~\ref{fig:TopoPhases}f,i). Secondly, the edge states below the lowest band show considerable magnetization on the edges with localization length around 3 unit cells and no magnetization in the bulk part of the chain.

Topological edge states in the band gap, on the other hand, show different properties for different parameter sets. As expected, a highly localized magnetization is obtained for parameters in Fig.~\ref{fig:TopoPhases}h, where the localization length is around two unit cells, but the same states for parameters in Fig.~\ref{fig:TopoPhases}e are almost totally delocalized. The reason lies in the behavior of the top levels in the lower band and bottom levels in the upper band before and after the quantum phase transition while increasing $\kappa_2$. In the disordered phase~I, the band structure is virtually the same as for the {\sc ssh} model, namely, top (bottom) levels of the bottom (top) band go down (up) in energy in phase~I$'$. Then the band gap closes at the topological phase boundary, and a pair of edge states separates from the bands. In the phase~I$''$, the top (bottom) levels of the bottom (top) band go in the opposite directions in energy, while the energy of the edge states remains constant (see Fig.~\ref{fig:TopoPhases}d up to $\kappa_2=0.5$).

In the symmetry broken phase~II, however, the behavior changes, as here the energy of both bands only grows with $\kappa_2$. If edge states have already appeared, the rate of growth of their energy is slow, compared to that of the top of the lower band, thus, eventually, these edge states get absorbed in the band, losing their spatial localization (around $\kappa_2=0.7$ in Fig.~\ref{fig:TopoPhases}d). If the quantum phase transition takes place before the topological phase transition (Fig.~\ref{fig:TopoPhases}c), the newly formed ``edge states'' get immediately absorbed in the lower band at the topological phase boundary and show no localization in magnetization  (Fig.~\ref{fig:TopoPhases}e).

\section{Conclusions}
\label{sec:conclusion}

We have discussed a one-dimensional lattice of coupled many-body systems with alternating couplings. Our model has the semiclassical features of the {\sc lmg} model combined with the topological character of the {\sc ssh} model. In the semiclassical limit $j\gg 1$, the whole lattice can be described at mean-field level. To obtain an arbitrary semiclassical trajectory, one needs to integrate the classical equations of motion for a system of $2N$ complex mean fields $\bm{\alpha}=(\alpha_{A1},\alpha_{B1},\ldots,\alpha_{AN},\alpha_{BN})$, where $N$ is the number of unit cells, and each unit cell contains two {\sc lmg} models. Although this situation is not problematic for small number $N$ of unit cells, the numerical integration of the equations of motion becomes prohibitive for large $N$. In this work, we were interested in ground-state properties. Therefore, we have focused on the global minima $\bm{\alpha}_m$ of the semiclassical energy landscape $E(\bm{\alpha})$~\cite{2011-Hayn-PRA,2014-Sorokin-PRE}. To calculate these minima, we have used the symmetries of the lattice to reduce the problem and to obtain analytic solutions in the case of a closed chain.

The analysis of the minima has revealed that the semiclassical energy landscape $E(\bm{\alpha})$ undergoes a bifurcation from a single minimum into a set of two degenerate minima. We have shown that quantum signatures of this bifurcation appear in the description of the quantum fluctuations. The semiclassical bifurcation is responsible for the existence of a second-order quantum phase transition of a symmetry breaking type. The phase diagram is depicted in Fig.~\ref{fig:Phases}a. However, the semiclassical bifurcation does not affect the whole critical behavior of the system. The geometry of the lattice itself can change the topological phase of the system without symmetry breaking. This kind of change of phase can appear within the symmetric phase, as well as in the symmetry-broken phase.

\acknowledgments{%
  The authors gratefully acknowledge financial support through DFG grants No.~BRA~1528/7,  No.~BRA~1528/8, No.~BRA~1528/9, No.~SFB~910 (G.E and T.B.), through Conselho Nacional de Desenvolvimento Científico e Tecnológico~-- Brazil (M.A.A.), through DAAD scholarship No.~91540276 (A.V.S.), through the National Research Foundation and Ministry of Education Singapore (partly through the Tier 3 Grant ``Random  numbers  from  quantum  processes''); and travel support by the EU IP-SIQS (V.M.B and D.G.A).%
 }

\appendix
\section{Coefficients of the quadratic Hamiltonian
\label{AppendixA}}
In this appendix, we show the explicit form of the coefficients appearing in the quadratic Hamiltonian Eq.~\eqref{eq:HQ}. The coefficients read

\begin{widetext}
 \begin{align*}
 	c^{(1)}(\alpha_{P},\alpha_{Q},\alpha_{R}) &= \Omega-\frac{\gamma}{8}\left[4-8|\alpha_P|^2-3\left(\alpha_P^2+(\alpha_P^*)^{2}\right)\right]
 	+\kappa_1\frac{\sqrt{2-|\alpha_Q|^2}(8-3|\alpha_P|^2)}{8(2-|\alpha_P|^2)^{3/2}}\left(\alpha_P^*\alpha_Q^{\phantom *}+\alpha_P^{\phantom *}\alpha_Q^*\right)\\
 	&\quad +\kappa_2\frac{\sqrt{2-|\alpha_R|^2}(8-3|\alpha_P|^2)}{8\left(2-|\alpha_P|^2\right)^{3/2}}\left(\alpha_P^*\alpha_R^{\phantom *}+\alpha_P^{\phantom *}\alpha_R^*\right), \\
 	c^{(2)}(\alpha_{P},\alpha_{Q})&= -\frac{16-12\left(|\alpha_P|^2+|\alpha_Q|^2\right)-9|\alpha_P|^2|\alpha_Q|^2+\alpha_P^2(\alpha_Q^*)^2}{8\sqrt{2-|\alpha_P|^2}\sqrt{2-|\alpha_Q|^2}}, \\
 	c^{(3)}(\alpha_{P},\alpha_{Q},\alpha_{R}) &= -\frac{\gamma}{8}\left[2-3|\alpha_P|^2-2(\alpha_P^*)^2\right]
 	+\kappa_1\frac{\alpha_P^*\sqrt{2-|\alpha_Q|^2}}{16(2-|\alpha_P|^2)^{3/2}}\left[(\alpha_P^*)^2\alpha_Q+\left(8-3|\alpha_P|^2\right)\alpha_Q^*\right] \\
 	&\quad +\kappa_2\frac{\alpha_P^*\sqrt{2-|\alpha_R|^2}}{16\left(2-|\alpha_P|^2\right)^{3/2}}\left[(\alpha_P^*)^2\alpha_R+\left(8-3|\alpha_P|^2\right)\alpha_R^*\right], \\
 	c^{(4)}(\alpha_{P},\alpha_{Q}) &=\frac{(\alpha_P^*)^2\left(4-3|\alpha_Q|^2\right)+(\alpha_Q^*)^2\left(4-3|\alpha_P|^2\right)}{8\sqrt{2-|\alpha_P|^2}\sqrt{2-|\alpha_Q|^2}}.
 	\label{eq:C}
 \end{align*}
 \end{widetext}

\bibliographystyle{apsrev4-1}
\bibliography{Xbib}

\end{document}